\documentclass[%
	aps,%
	prl,%
	twocolumn,%
	10pt,%
	balancelastpage,%
	amsmath,%
	amssymb,%
	superscriptaddress,%
	floatfix,%
	footinbib]{revtex4-2}

\usepackage[utf8]{inputenc}
\usepackage[T1]{fontenc}
\usepackage{xcolor}
\usepackage{graphicx}
\usepackage{braket}
\usepackage{orcidlink}
\usepackage{hyperref}
\hypersetup{colorlinks=true,allcolors=blue,pageanchor=true}

\newcommand{\mytitle}{Range-controlled entanglement in Lindbladian skin states of monitored fermions}

\begin{document}
	\title{\mytitle}
	
	\author{Gianluca Passarelli\,\orcidlink{0000-0002-3292-0034}}
	\email{gianluca.passarelli@unina.it}
	\author{Angelo Russomanno\,\orcidlink{0009-0000-1923-370X}}
	\email{angelo.russomanno@unina.it}
	\affiliation{Dipartimento di Fisica, Universit\`a di Napoli ``Federico II'', I-80126 Napoli, Italy}
	
	\author{Davide Rossini\,\orcidlink{0000-0002-9222-1913}}
	\affiliation{Dipartimento di Fisica dell’Università di Pisa and INFN, Largo Pontecorvo 3, I-56127 Pisa, Italy}
	
	\author{Procolo Lucignano\,\orcidlink{0000-0003-2784-8485}}
	\affiliation{Dipartimento di Fisica, Universit\`a di Napoli ``Federico II'', I-80126 Napoli, Italy}

	\begin{abstract}
		Reservoir engineering can stabilize states inaccessible to unitary dynamics. Directed particle-conserving dissipation creates Lindbladian skin states, where Pauli exclusion turns edge accumulation into a many-body density imbalance. In a monitored fermion chain with tunable hopping range, we identify, within a Gaussian trajectory approximation, two finite-size scaling regimes: short-range hopping is consistent with complete skin accumulation and area-law entanglement, whereas sufficiently long-range hopping produces a finite bulk tail and effective algebraic sub-volume-law entanglement. Dissipation and coherent hopping thus jointly control skin localization and quantum entanglement, highlighting their close interconnection.
	\end{abstract}
	
	\maketitle
	
	\textit{Introduction ---} Reservoir engineering and continuous monitoring can shape the quantum-information structure of many-body steady states. Non-Hermitian Hamiltonians provide an effective description of open quantum systems and can host unconventional phenomena such as exceptional points, $\mathcal{PT}$-symmetry breaking, and anomalous topological behavior~\cite{Ashida_2020,Rut,Martinez_Alvarez_2018,Wang_2021,RevModPhys.93.015005,Ding_2022,PhysRevLett.121.086803}. Among their most striking manifestations is the non-Hermitian skin effect, in which an extensive number of eigenstates become exponentially localized at the boundaries of an open system~\cite{Zhang31122022}. In the paradigmatic Hatano-Nelson chain~\cite{PhysRevLett.77.570,PhysRevB.56.8651}, nonreciprocal hopping piles particles at an edge; for fermions, Pauli exclusion converts this into a many-body density imbalance and a Fermi-skin profile~\cite{shen2025observationnhse}. Skin accumulation has by now been observed in photonic quantum walks and ultracold-atom platforms~\cite{Xiao2020NatPhys,Liang2022PRL,Zhao2025Nature}, highlighting the experimental relevance of dissipative and nonunitary routes to non-Hermitian physics.
	
	This naturally raises the question of how such physics emerges in a genuinely dissipative many-body setting, where the dynamics is governed by a Lindblad equation rather than by an effective non-Hermitian Hamiltonian. Open-system skin physics has been studied through Liouvillian spectra and chiral damping~\cite{Song2019PRL}, generic quadratic Lindbladians~\cite{ZhouYu2022PRA}, and interacting dissipative settings~\cite{Hu2025PRL,qrh6-vx64}. It has also emerged as an entanglement mechanism in no-click non-Hermitian dynamics, where skin effects can drive sharp entanglement transitions~\cite{PhysRevX.13.021007,li2023disorderinducedentanglementphasetransitions,Gal}. These results connect naturally to monitored dynamics, where quantum trajectories capture the competition between entanglement growth and measurement-induced purification~\cite{PhysRevX.10.041020, Li2018, Chan2019, Skinner2019, Szyniszewski2019, Vasseur2021, Bao2021, Nahum2020, Chen2020,Li2019, Jian2020, Li2021, Szyniszewski2020, Turkeshi2020, Lunt2021, Sierant2022_B, Nahum2021, Zabalo2020, Sierant2022_A, Chiriaco2023, Klocke2023,lirasolanilla2024,Nehra_2025, PhysRevB.109.214204, PhysRevResearch.6.043246, chahine2023entanglement, delmonte2024, Passarelli_2024, DeLuca2019, Buchhold2021,Jian2022, Coppola2022, Fava2023, Poboiko2023, Jian2023, Merritt2023, Alberton2021, Turkeshi2021, Szyniszewski2022, Turkeshi2022, Piccitto2022, Piccitto2022e, Piccitto2023, Tirrito2022, Paviglianiti2023, Lang2020, Minato2022, Zerba2023, paviglianiti2023enhanced,chatterjee2024, gda_EPJB, Le_Gal_2024,3sv9-2tkw,Russomanno2023_longrange,Lunt2020,Rossini2020, Tang2020, Fuji2020, Sierant2021, Doggen2022, Altland2022, li2024monitored,Ippoliti2021, Sriram2022}. The non-Hermitian settings of Refs.~\cite{PhysRevX.13.021007,li2023disorderinducedentanglementphasetransitions,Gal} correspond to the no-click limit~\cite{Turkeshi2021}, where the environment is monitored but no measurement events occur.

	Here, instead, we address the full jump-resolved Lindblad dynamics. We study a particle-conserving Lindbladian whose asymmetric quadratic jump operators generate a Lindbladian skin effect (LSE)~\cite{PhysRevX.13.021007,PhysRevLett.127.070402}. The dynamics is unraveled into quantum-jump trajectories and use a Gaussian trajectory approximation based on normal two-point correlators~\footnote{See Supplemental Material, where we derive the equations of motion in the Gaussian approximation, benchmark them against exact diagonalization, and provide additional data complementing the discussion in the main text.}. In the exact unraveling, averaging over trajectories reproduces Lindblad expectation values of linear observables. Within the Gaussian closure, this correspondence is approximate and is benchmarked below~\cite{Note1}. Nonlinear quantities, such as entanglement entropies, are instead evaluated for each trajectory and averaged afterwards.
	
	\begin{figure}[tb]
		\centering
		\includegraphics[width=\columnwidth]{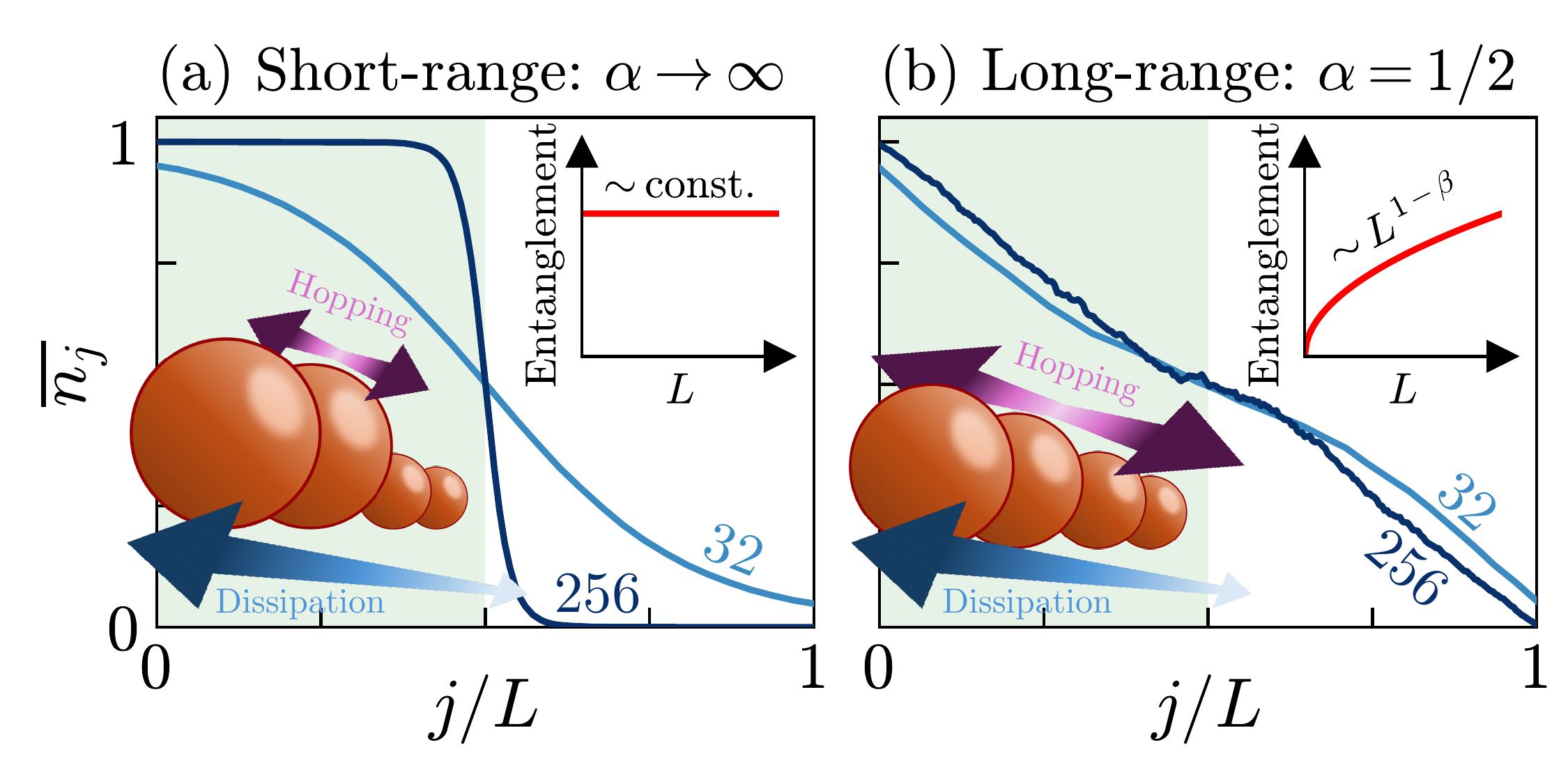}
		\caption{Representative steady-state fermion density profiles at half filling for $\gamma = 1/2$ and $\lambda = 2$. The shaded region marks the left half of the chain, where directed dissipation accumulates particles. (a) Short-range hopping sharpens the Fermi-skin profile with increasing $L$ (see curve labels), consistent with complete skin accumulation; the inset sketches the corresponding area-law behavior of the trajectory entanglement entropy. (b) Long-range hopping leaves a finite density tail in the right half of the chain over the accessible sizes, indicating incomplete skin accumulation; the inset sketches the corresponding sub-volume-law finite-size growth of entanglement.}
		\label{fig:sketch}
	\end{figure}

    We then introduce number-conserving coherent hopping with tunable range at half filling. The hopping range controls two closely related properties of the monitored skin state. Short-range hopping drives the system toward a Pauli-blocked, complete skin profile with area-law Gaussian-trajectory entanglement. By contrast, sufficiently long-range hopping leaves a finite bulk tail and yields entanglement growth consistent with an algebraic sub-volume-law over the accessible system sizes (Fig.~\ref{fig:sketch}). Thus, beyond the no-click limit, the full jump-resolved Lindblad dynamics reveals a range-controlled correspondence between density accumulation and monitored many-body entanglement. We remark that the skin effect is crucial: Inhibiting it using periodic boundary conditions or a balanced Lindbladian transforms the area law of the short-range imbalanced case into a size-independent crossover in the Hamiltonian coupling between area-law and power-law behavior.
    
	\textit{Model ---}	We first specify the dissipative part of the Lindbladian for spinless fermions on an open chain of $L$ sites~\cite{PhysRevX.13.021007,PhysRevLett.127.070402}. Its particle-conserving jump operators are the quantum analogue of biased exclusion dynamics~\cite{Temme_2012} and generate a dissipative skin effect when left-right symmetry is broken~\cite{Song2019PRL,ZhouYu2022PRA}:
	\begin{equation}
		\label{eq:lindblad}
		\mathcal{L}\hat{\rho}
		=
		\sum_{\ell=1}^{L-1}\sum_{w=R,L}
		\left(
		\hat{L}_{\ell w}\hat{\rho}\hat{L}_{\ell w}^\dagger
		-\frac{1}{2}\left\{\hat{L}_{\ell w}^\dagger\hat{L}_{\ell w},\hat{\rho}\right\}
		\right),
	\end{equation}
	with Lindblad operators
	\begin{equation}
			\hat{L}_{\ell R}=\sqrt{\frac{J-\gamma}{2}}\,\hat{c}_{\ell+1}^\dagger \hat{c}_\ell,\quad 
			\hat{L}_{\ell L}=\sqrt{\frac{J+\gamma}{2}}\,\hat{c}_{\ell}^\dagger \hat{c}_{\ell+1}.
	\end{equation}
	Here $J>0$ sets the overall dissipative scale and $\gamma$ controls the left-right asymmetry, with $0\le \lvert\gamma\rvert\le J$ so that all jump rates are non-negative. For $\gamma=0$ the dissipation is symmetric and does not sustain a skin effect, while for $\gamma=\pm J$ one direction is fully suppressed. In the following we restrict to $\gamma>0$ without loss of generality and set $J=1$ as the energy scale.
		
	Without unitary driving, the dynamics is classical in the occupation basis: trajectories remain product Fock states with identically vanishing entanglement, while the diagonal steady-state ensemble displays the Lindbladian skin effect for any $\gamma\neq0$. Adding coherent hopping makes the problem genuinely many-body: the symmetric unitary drive competes with the asymmetric dissipation and can qualitatively alter both the skin accumulation and the entanglement of the steady state. To probe this interplay, we supplement the dissipative dynamics with the Hamiltonian
	\begin{equation}
		\label{eq:hamiltonian-power-law}
		\hat{H}_\alpha
		=
		\frac{\lambda}{2 \mathcal{K}_{\alpha}}
		\sum_{j=1}^{L-1}\sum_{r=1}^{L-j}
		\frac{1}{r^\alpha}
		\left(
		\hat{c}_{j+r}^\dagger \hat{c}_j
		+
		\hat{c}_{j}^\dagger \hat{c}_{j+r}
		\right).
	\end{equation}
	The complete evolution is governed by the master equation $\partial_t\hat\rho=-i[\hat H_\alpha,\hat\rho]+\mathcal L\hat\rho \quad (\hbar = 1)$. Open boundary conditions are always assumed. The Kac scaling factor $\mathcal{K}_\alpha = (1/L) \sum_{r=1}^{L-1} (L - r)/r^\alpha$ ensures that the Hamiltonian~\eqref{eq:hamiltonian-power-law} is extensive for all $\alpha$~\cite{10.1063.1.1703946}. The exponent $\alpha$ controls the hopping range: in one dimension, the coherent model is effectively infinite-range for $\alpha<1$, long-range for $1<\alpha<2$, and effectively short-range for $\alpha>2$~\cite{PhysRevX.10.031009,PhysRevLett.130.070401}. In the dissipative skin problem, $\alpha\simeq1$ marks a point with maximum entanglement growth, while the recovery of area-law entanglement occurs inside $1<\alpha<2$. Representative steady-state density profiles are shown in Fig.~\ref{fig:sketch}.
	
	We seek to relate the steady-state entanglement of this model to the Lindbladian skin effect. To this end, we unravel the Lindblad evolution into quantum-jump trajectories~\cite{Plenio,Daley2014}, as detailed in the Supplemental Material~\cite{Note1}. Since the no-jump generator contains quartic density-density terms, the resulting trajectory dynamics is non-Gaussian. To access large system sizes, we therefore employ a Gaussian trajectory approximation: four- and six-point correlators are Wick-factorized and the normal correlation matrix $\mathcal{G}_{mn}=\langle \hat{c}_m^\dagger \hat{c}_n\rangle$ is evolved directly. Since the particle number $N$ is conserved and the initial state has fixed $N$, anomalous correlators vanish identically. Benchmarks against exact trajectory simulations are provided in the Supplemental Material~\cite{Note1}.

	To probe the presence of the LSE, we consider the half-chain imbalance operator
	\begin{equation}
		\hat{I}
		=
		\frac{1}{N}
		\left(
		\sum_{j>L/2}\hat{n}_j
		-
		\sum_{j\leq L/2}\hat{n}_j
		\right).
	\end{equation}
    We evaluate this quantity along each trajectory and then average over the ensemble, thereby reconstructing the Lindblad evolution. We focus on the steady-state behavior at long times and denote the steady-state value by $\overline{I}$. This intensive observable vanishes for an even density profile and reaches $\lvert\overline I\rvert=1$ when all particles occupy one half of the chain. At half filling and for $\gamma>0$, we call the LSE \emph{complete} if $\overline I_\infty\equiv\lim_{L\to\infty}\overline I=-1$, and \emph{incomplete} if $-1<\overline I_\infty<0$.
	
	We also study the half-chain entanglement entropy. Since the Gaussian approximation evolves only the two-point correlators, the trajectory entanglement is evaluated from the restriction $\mathcal{G}_A$ of the correlation matrix to subsystem $A$ --- the first half of the chain. Denoting by $\{\nu_a\}$ the eigenvalues of $\mathcal{G}_A$, we define
	\begin{equation}
		\label{eq:gaussian-entropy}
		S_A^{(\mathrm{G})}
		=
		-\sum_a\left[\nu_a\log\nu_a+(1-\nu_a)\log(1-\nu_a)\right].
	\end{equation}
	For exact Gaussian states this expression coincides with the von Neumann entanglement entropy; here it should be understood as the trajectory-wise entropy of the Gaussian surrogate state associated with the approximate correlation matrix. In the following, we denote by $\overline{S_{L/2}}$ the steady-state average of this quantity over time and over stochastic realizations after the transient regime. Its reliability is benchmarked in Ref.~\cite{Note1} against exact simulations at small sizes.

    \begin{figure}[tb]
		\centering
		\includegraphics[width = \columnwidth]{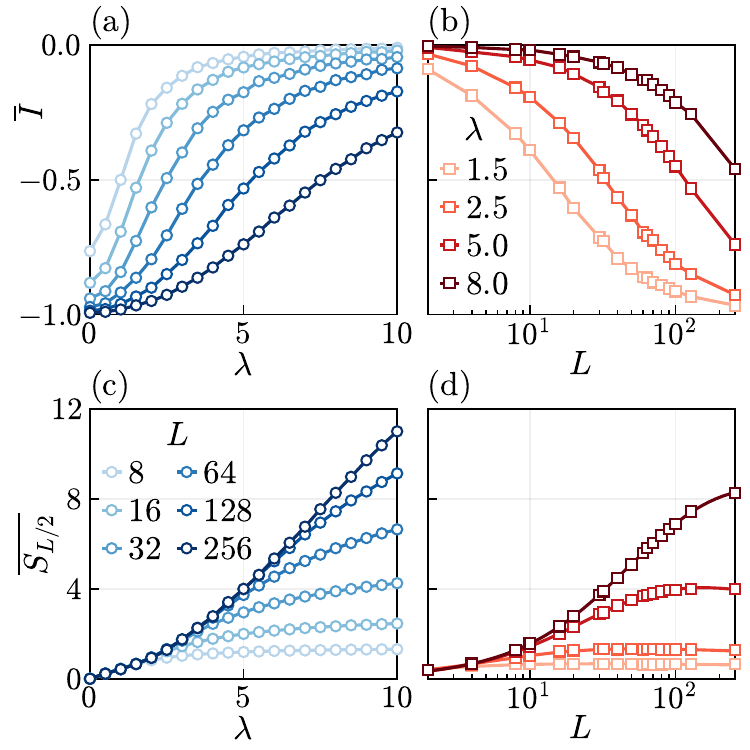}
        \caption{Data for the nearest-neighbor Hamiltonian ($\gamma=1/2$): (a) imbalance $\overline I$ and (c) half-chain entanglement entropy $\overline{S_{L/2}}$ as functions of the coherent coupling $\lambda$. (b) Imbalance and (d) half-chain entanglement entropy as functions of the system size $L$. In (d), solid lines show fits to Eq.~\eqref{eq:entanglement-fit}. Within the reliable fitting window, they yield $\beta \ge 1$, consistent with the absence of algebraic growth.}
	   \label{fig:gamma-0.5-nn}
	\end{figure}
    
	\textit{Results ---} We fix $\gamma=1/2$ as a representative asymmetric point, even though results do not qualitatively depend on this choice. Data are obtained from $N_T=100$ trajectories evolved up to $T=1000$, with steady-state averages over $t\in[900,1000]$; statistical fluctuations are smaller than the symbol sizes. Since the Gaussian dynamics has a unique steady state in each charge sector, we initialize a N\'eel state at half filling $\ket{\psi} = \prod_{i=1}^N \hat{c}_{2i-1}^\dagger \ket{\emptyset}$, with $N=L/2$, where $\ket{\emptyset}$ is the fermionic vacuum.
	
	\textit{Short range.} We begin with the nearest-neighbor limit (i.e., $\alpha\to\infty$) of the Hamiltonian~\eqref{eq:hamiltonian-power-law}, where only terms with $r = 1$ survive.
	Figure~\ref{fig:gamma-0.5-nn}(a) shows the steady-state imbalance as a function of the coherent coupling $\lambda$. Small $\lambda$ corresponds to a strongly dissipative regime, while large $\lambda$ is dominated by coherent dynamics. At $\lambda=0$, where the evolution is purely dissipative, the imbalance approaches $\overline I=-1$ with increasing $L$, indicating a complete Lindbladian skin effect in the thermodynamic limit. For finite values of $\lambda$, Fig.~\ref{fig:gamma-0.5-nn}(b) shows the same flow toward the fully imbalanced state. The balanced limit is only recovered for $\lambda\to\infty$, i.\,e., where dissipation becomes negligible. The density profiles shown in Fig.~\ref{fig:sketch}(a) confirm that the particle density in the right half of the chain decreases with increasing $L$. Our data therefore suggest a complete LSE throughout the short-range model for any finite ratio between coherent and dissipative couplings.
	
	\begin{figure}[tb]
		\centering
		\includegraphics[width = \columnwidth]{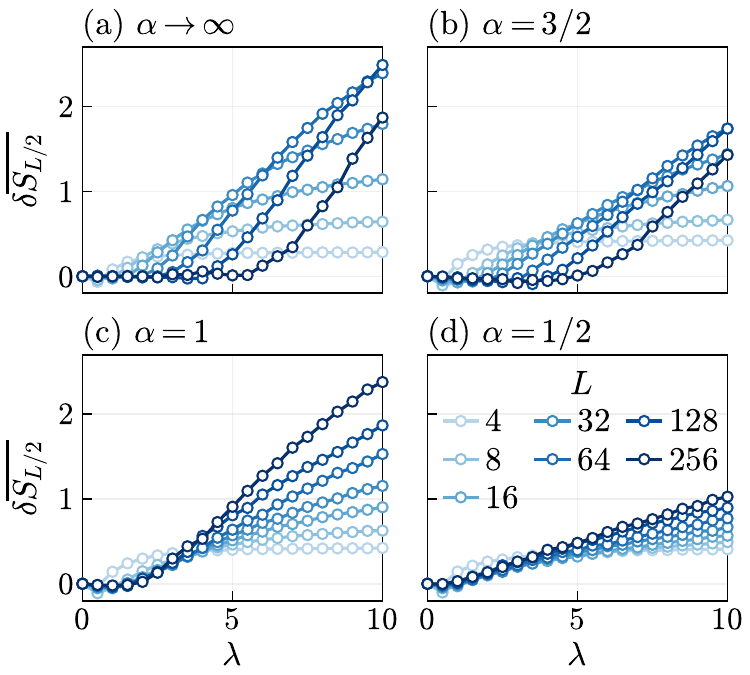}
		\caption{Finite-size difference $\delta \overline{S_{L/2}}$ for {(a)}~the nearest-neighbor model ($\alpha\to\infty$), {(b)}~$\alpha = 3/2$, {(c)}~$\alpha = 1$, and {(d)}~$\alpha = 1/2$. The onset of entanglement growth drifts toward larger $\lambda$ with increasing $L$ for $\alpha\to\infty$ and, over the accessible sizes, also for $\alpha=3/2$, while it remains approximately size-independent for $\alpha=1$ and $\alpha=1/2$.}
		\label{fig:gamma-0.5-nn-delta-entanglement}
	\end{figure}
	
	\begin{figure*}[tb]
		\centering
		\includegraphics[width = \linewidth]{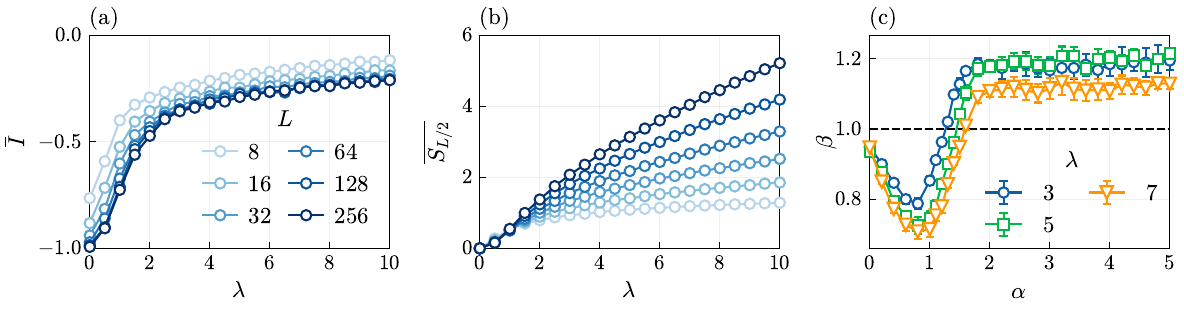}
		\caption{(a,b) Data for the long-range Hamiltonian with $\alpha=1/2$ ($\gamma=1/2$): (a) imbalance $\overline I$ and (b) half-chain entanglement entropy $\overline{S_{L/2}}$ as functions of the coherent coupling $\lambda$. (c) Fitting exponent $\beta$ [Eq.~\eqref{eq:entanglement-fit}] as a function of the hopping range $\alpha$ for representative values of $\lambda$. The dashed line marks $\beta=1$. The minimum near $\alpha\simeq1$ indicates maximal sub-volume-law growth, while the recovery of $\beta\ge1$ occurs within the interval $1 \lesssim \alpha \lesssim 2$.}
		\label{fig:gamma-0.5-alpha-0.5}
	\end{figure*}
	
	We now turn to the entanglement entropy averaged over the trajectory ensemble in the steady-state regime. In Fig.~\ref{fig:gamma-0.5-nn}(c), the curves collapse at small $\lambda$, consistent with an area-law regime, whereas at larger $\lambda$ the entropy increases over the accessible system sizes. This behavior is shown more explicitly in Fig.~\ref{fig:gamma-0.5-nn}(d). %
    To quantify the nature of the entanglement scaling, we fit the data with the empirical form
	\begin{equation}
		\label{eq:entanglement-fit}
		S(L)=\frac{A L}{1+C L^\beta}, \qquad A,C,\beta\ge 0,
	\end{equation}
	introduced in Ref.~\cite{3sv9-2tkw} to capture a crossover from small-size linear growth to large-size sublinear scaling in monitored integrable fermionic systems. We use this {\em ansatz} as an empirical diagnostic: $\beta=0$ indicates volume-law scaling, $0<\beta<1$ sub-volume-law growth, and $\beta\ge1$ absence of algebraic growth, with possible logarithmic corrections at $\beta=1$~\cite{Note1}. The fits accurately describe the data and outperform alternative forms tested in the Supplemental Material~\cite{Note1}. For the accessible sizes, the nearest-neighbor data yield effective $\beta>1$, consistent with area-law behavior.
	
	The same conclusion is reinforced by the finite-size difference
	\begin{equation}
		\label{eq:dS}
		\delta \overline{S_{L/2}}=\overline{S_{L/2}(L)}-\overline{S_{L/4}(L/2)},
	\end{equation}
	namely the difference between the half-chain entanglement entropies for systems of size $L$ and $L/2$, respectively. For an area law, $\delta \overline{S_{L/2}}\to0$; for logarithmic scaling, it tends to a constant; for power-law growth, it increases with $L$. As shown in Fig.~\ref{fig:gamma-0.5-nn-delta-entanglement}(a), $\delta \overline{S_{L/2}}$ is already negligible for small $\lambda$. At larger $\lambda$, a growth with the system size emerges, but the crossover shifts to higher values of $\lambda$ as $L$ increases. This drift supports the interpretation that the observed enhancement of entanglement reflects a finite-size crossover and that the short-range regime remains consistent with area-law at finite $\lambda$, within the Gaussian approximation.
	We therefore interpret the short-range data as consistent with complete Lindbladian skin accumulation together with area-law steady-state entanglement at finite coherent coupling.
    
    \textit{Long range.} The above scenario changes qualitatively as the hopping range increases. To illustrate this point, we now consider the long-range Hamiltonian with $\alpha=1/2$.
    Figure~\ref{fig:gamma-0.5-alpha-0.5}(a) shows the steady-state imbalance as a function of the coherent coupling $\lambda$. In contrast to Fig.~\ref{fig:gamma-0.5-nn}(a), increasing $L$ does not drive the data toward $\overline I=-1$; instead, the curves approach a nonmaximal limiting value~\cite{Note1}. As shown in Fig.~\ref{fig:sketch}(b), the long-range hopping causes particles to drift to the right half of the chain for all sizes, despite asymmetric incoherent processes pushing them to the left. Our data are therefore consistent with an incomplete Lindbladian skin effect in the thermodynamic limit.
    
    The entanglement structure changes sharply as well. As shown in Fig.~\ref{fig:gamma-0.5-alpha-0.5}(b), the steady-state entropy is approximately size-independent for $\lambda<1$ and grows with $L$ for $\lambda>1$, within the Gaussian approximation. Fits to Eq.~\eqref{eq:entanglement-fit} yield $\beta<1$, consistent with algebraic sub-volume-law growth, while the finite-size difference $\delta \overline{S_{L/2}}$ exhibits a size-independent crossover near $\lambda_c\simeq1$ [Fig.~\ref{fig:gamma-0.5-nn-delta-entanglement}(d)].

	The dependence on the hopping range is more nuanced than a sharp boundary. As shown in Fig.~\ref{fig:gamma-0.5-alpha-0.5}(c), the effective finite-size exponent $\beta(\alpha)$ is minimal close to $\alpha\simeq1$, where the observed sub-volume-law growth is most pronounced, while the empirical transition to $\beta\ge1$ occurs in the range $1 \lesssim \alpha \lesssim 2$ and depends only weakly on $\lambda$. Thus $\alpha\simeq1$ corresponds to maximal entanglement generation in our finite-size Gaussian data, whereas $\alpha=3/2$ lies close to the crossover toward area-law behavior, where logarithmic corrections may play a significant role~\cite{Note1}. For $\alpha>2$ the exponent reaches a plateau and attains the same value as the  nearest-neighbor ($\alpha\to\infty$) limit.
    
    The crossover is controlled by a long-distance balance between coherent driving and dissipation and can be intuitively understood from the following argument. Dissipative jumps generate a local drift toward the left boundary, while Pauli exclusion packs the half-filled gas into a skin domain. Particles deep in the filled region, and holes deep in the empty region, are inert. The relevant entangling channels therefore originate from the active interfacial/tail region where the local occupation fluctuates. In the Supplemental Material~\cite{Note1} we estimate the width of the interface of the Fermi skin profile. In the short-range regime, we observe a sharp interface, consistent with area-law entanglement scaling. In contrast, in the long-range regime an extended skin develops, leading to a macroscopic number of active sites at the interface.
    
    We emphasize that the existence of this skin domain is crucial: periodic boundary conditions or left-right symmetric dissipation ($\gamma=0$) qualitatively alter this scenario, yielding a transition between area-law behavior and algebraically growing entanglement even for short-range hopping~\cite{Note1}. The Hamiltonian hopping range thus provides a direct way to engineer distinct Lindbladian skin-effect regimes with qualitatively different quantum-state properties. 
	
	\textit{Conclusions ---} We have shown that directed particle-conserving dissipation and coherent hopping range jointly control Lindbladian skin accumulation and monitored entanglement. Our quantum-jump simulations, performed within a Gaussian trajectory approximation benchmarked at small sizes, reveal two distinct regimes.
    Short-range hopping produces a complete Pauli-blocked skin state with area-law entanglement. By contrast, sufficiently long-range hopping couples the skin domain to the bulk, leading to incomplete skin accumulation and algebraically growing sub-volume-law entanglement. The strongest growth occurs near $\alpha\simeq1$, while the recovery of area-law behavior takes place in the range $1 \lesssim \alpha \lesssim 2$. Control simulations without a skin domain confirm that this effect relies on the interplay between directed dissipation, Pauli blocking, and coherent hopping range.

    The ingredients are compatible with digital simulations of nonunitary lattice dynamics realizing Fermi-skin physics~\cite{shen2025observationnhse}, circuit implementations of Lindblad evolution~\cite{Han_2021,Hu_2020}, and variational simulations of dissipative many-body dynamics~\cite{liu2025variationalquantumsimulationmanybody}. They also connect naturally to cold-atom routes to engineered dissipation~\cite{PhysRevLett.105.227001,Yi_2012} and to ultracold-atom realizations of non-Hermitian physics~\cite{Liang2022PRL,Zhao2025Nature}. Until now Gaussian approximations of the dynamics along trajectories have only been used for bosonic models~\cite{Verstraelen_2018,Verstraelen_2019,PhysRevResearch.2.022037}, we apply them here to a fermionic case. Density profiles and imbalance provide the most direct experimental probes, whereas entanglement requires access to the monitored conditional ensemble.
    
    \begin{acknowledgments}
    	\textit{Acknowledgments ---} We acknowledge valuable discussions with G.\,Salatino. P.\,L., G.\,P., and A.\,R. acknowledge support from PNRR MUR Project PE0000023-NQSTI.
    \end{acknowledgments}

	\clearpage
	\widetext
	\makeatletter
	\begin{center}
		\textbf{\large \mytitle\\[1ex] Supplemental Material}
	\end{center}
	\makeatother
	\setcounter{equation}{0}
	\setcounter{figure}{0}
	\setcounter{table}{0}
	\setcounter{page}{1}
	\makeatletter
	\renewcommand{\theequation}{S\arabic{equation}}
	\renewcommand{\thefigure}{S\arabic{figure}}
	\renewcommand{\thesection}{S\arabic{section}}
	\renewcommand{\bibnumfmt}[1]{[S#1]}
	\renewcommand{\citenumfont}[1]{#1}
    \renewcommand{\thepage}{S\arabic{page}}
	
	In this Supplemental Material, we derive the quantum jump equations of motion for the correlation matrix in the Gaussian approximation. We compare the results obtained with these equations to exact diagonalization results. Moreover, we show additional data to complement the discussion of the main text.
	
	\section{Quantum-jump equations and the Gaussian approximation}
	\label{app:quantum-jumps-gaussian}
	
	To study the dynamics of entanglement, we unravel the Lindblad master equation $\partial_t \hat \rho = -i [\hat H_\alpha, \hat \rho] + {\mathcal{L}} \hat \rho$, with ${\mathcal{L}} \hat \rho$ given in Eq.~\eqref{eq:lindblad} and $\hat H_\alpha$ in Eq.~\eqref{eq:hamiltonian-power-law} of the main text, into stochastic quantum trajectories described by quantum jumps. For the sake of clarity, in this section we keep $J$ explicit and do not set $J=1$ as the energy scale. Upon discretizing time in steps $\delta t$, the state is updated according to the following rules:
	\begin{itemize}
		\item For a given bond $\ell$, with probability
		\begin{equation}
			\label{pelle:eqn}
				p_\ell(t)
				=
				\delta t\,\braket{\psi_t|\hat{L}_{\ell R}^\dagger \hat{L}_{\ell R}|\psi_t}
				=
				\delta t\left(\frac{J-\gamma}{2}\right)
				\braket{\psi_t|\hat{c}_\ell^\dagger \hat{c}_{\ell+1}\hat{c}_{\ell+1}^\dagger \hat{c}_{\ell}|\psi_t},
		\end{equation}
		a right jump occurs and the state is updated into
		\begin{equation}
			\label{jumpelle:eqn}
			\ket{\psi_t}\to\ket{\psi_{t+\delta t}}
			=
			\frac{\hat{L}_{\ell R}\ket{\psi_t}}
			{\lVert \hat{L}_{\ell R}\ket{\psi_t}\rVert}
			=
			\frac{\hat{c}_{\ell+1}^\dagger \hat{c}_{\ell}\ket{\psi_t}}
			{\lVert \hat{c}_{\ell+1}^\dagger \hat{c}_{\ell}\ket{\psi_t}\rVert}.
		\end{equation}
		
		\item For a given bond $\ell$, with probability
		\begin{equation}
				q_\ell(t)
				=
				\delta t\,\braket{\psi_t|\hat{L}_{\ell L}^\dagger \hat{L}_{\ell L}|\psi_t}=
				\delta t\left(\frac{J+\gamma}{2}\right)
				\braket{\psi_t|\hat{c}_{\ell+1}^\dagger \hat{c}_{\ell}\hat{c}_{\ell}^\dagger \hat{c}_{\ell+1}|\psi_t},
		\end{equation}
		a left jump occurs and the state is updated into
		\begin{equation}
			\label{jumperre:eqn}
			\ket{\psi_t}\to\ket{\psi_{t+\delta t}}
			=
			\frac{\hat{L}_{\ell L}\ket{\psi_t}}
			{\lVert \hat{L}_{\ell L}\ket{\psi_t}\rVert}
			=
			\frac{\hat{c}_{\ell}^\dagger \hat{c}_{\ell+1}\ket{\psi_t}}
			{\lVert \hat{c}_{\ell}^\dagger \hat{c}_{\ell+1}\ket{\psi_t}\rVert}.
		\end{equation}
		
		\item With the remaining probability
		\begin{equation}
			p(t)=1-\sum_{\ell=1}^{L-1}\left[p_\ell(t)+q_\ell(t)\right],
		\end{equation}
		the state evolves under the effective non-Hermitian Hamiltonian
        \begin{equation}
				\hat{H}_\mathrm{eff}
				\equiv
				\hat{H}_\alpha
				-\frac{i}{2}\sum_{\ell=1}^{L-1}
				\left(
				\hat{L}_{\ell R}^\dagger \hat{L}_{\ell R}
				+
				\hat{L}_{\ell L}^\dagger \hat{L}_{\ell L}
				\right)
				=
				\hat{H}_\alpha
				-i\sum_{\ell=1}^{L-1}
				\left[
				\left(\frac{J-\gamma}{4}\right)\hat{n}_\ell
				+
				\left(\frac{J+\gamma}{4}\right)\hat{n}_{\ell+1}
				-
				\frac{J}{2}\hat{n}_\ell \hat{n}_{\ell+1}
				\right],
		\end{equation}
        in such a way that
		\begin{equation}
			\label{nH:eqn}
			\ket{\psi_t} \to \ket{\psi_{t+\delta t}}
			=
			\frac{e^{-i\delta t \hat{H}_\mathrm{eff}}\ket{\psi_t}}
			{\lVert e^{-i\delta t \hat{H}_\mathrm{eff}}\ket{\psi_t}\rVert}.
		\end{equation}
	\end{itemize}
	
	The dynamics conserves the total particle number $N$, and throughout we focus on half filling, $N=L/2$. Since $\hat{H}_\mathrm{eff}$ contains quartic terms, the state along a given trajectory is generally non-Gaussian. To access larger system sizes, we therefore adopt a Gaussian approximation in which four- and six-point correlators are factorized into products of two-point correlators through Wick's theorem and fermionic anticommutation relations. This allows us to simulate the quantum-jump dynamics directly at the level of the normal correlation matrix
	\begin{equation}
		\mathcal{G}_{mn}=\langle \hat{c}_m^\dagger \hat{c}_n\rangle,
	\end{equation}
	within a waiting-time-distribution (WTD) algorithm. Anomalous correlators remain identically zero as a consequence of particle-number conservation and of the fact that we always choose initial states with a fixed number of fermions. Below, we benchmark this approximation with exact trajectory simulations at small sizes and show that it captures the qualitative behavior of both steady-state observables and entanglement.
	
	In the WTD algorithm, the system is initialized in a normalized state $\ket{\psi(0)}$ and a random number $r$, uniformly distributed in $[0,1]$, is drawn. The state is then evolved under $\hat{H}_\mathrm{eff}$, yielding an unnormalized state $\ket{\tilde{\psi}(t)}$. This deterministic evolution proceeds until the squared norm
	\begin{equation}
		\mathcal{N}(t)=\langle \tilde{\psi}(t)|\tilde{\psi}(t)\rangle
	\end{equation}
	reaches the value $r$. The corresponding time $t_J$ identifies a quantum jump. The jump channel $(j,w)$ is then selected with probability proportional to $\langle \hat{L}_{j,w}^\dagger \hat{L}_{j,w}\rangle$ evaluated at $t_J$, the state is renormalized after the jump, a new random number is drawn, and the procedure is iterated.
	
	The first ingredient of the WTD scheme is therefore the equation of motion for $\mathcal{N}(t)$,
	\begin{equation}
		\dot{\mathcal{N}}(t)
		=
		-\mathcal{N}(t)\sum_{j=1}^{L-1}\sum_{w=L,R}
		\langle \hat{L}_{j,w}^\dagger \hat{L}_{j,w}\rangle
		=
		-\Lambda(t) \, \mathcal{N}(t),
	\end{equation}
	with initial condition $\mathcal{N}(0)=1$. For the present choice of jump operators, $\Lambda(t)$ depends on four-point correlators. Within the Gaussian approximation one finds
	\begin{equation}
			\Lambda(t)
			\approx
			\sum_{j=1}^{L-1}\Bigg[
			\frac{J-\gamma}{2}
			\Big(
			\mathcal{G}_{jj}
			-\mathcal{G}_{jj} \,\mathcal{G}_{j+1,j+1}
			+\mathcal{G}_{j,j+1} \,\mathcal{G}_{j+1,j}
			\Big)
			+\frac{J+\gamma}{2}
			\Big(
			\mathcal{G}_{j+1,j+1}
			-\mathcal{G}_{jj} \, \mathcal{G}_{j+1,j+1}
			+\mathcal{G}_{j,j+1} \, \mathcal{G}_{j+1,j}
			\Big)
			\Bigg]\,,
	\end{equation}
    where the symbol ``$\approx$'' marks the Gaussian approximation. The update rule for the correlation matrix after a right jump $\hat{L}_{j,R}$ takes the form
	\begin{equation}
		\mathcal{G}_{mn}\to \frac{N_R}{D_R},
	\end{equation}
	where
		\begin{equation}
			\begin{aligned}
				N_R
				&=
				\langle
				\hat{c}_j^\dagger \hat{c}_{j+1}
				\hat{c}_m^\dagger \hat{c}_n
				\hat{c}_{j+1}^\dagger \hat{c}_j
				\rangle \\
				&\approx
				-\mathcal{G}_{j,j}\mathcal{G}_{j+1,j+1}\mathcal{G}_{m,n}
				+\mathcal{G}_{j,j}\mathcal{G}_{j+1,n}\mathcal{G}_{m,j+1}
				-\mathcal{G}_{j,j}\mathcal{G}_{j+1,n}\,\delta_{j+1,m}
				-\mathcal{G}_{j,j}\mathcal{G}_{m,j+1}\,\delta_{j+1,n}
				+\mathcal{G}_{j,j}\mathcal{G}_{m,n}\\
				&\quad
				+\mathcal{G}_{j,j}\delta_{j+1,m}\delta_{j+1,n}
				+\mathcal{G}_{j,j+1}\mathcal{G}_{j+1,j}\mathcal{G}_{m,n}
				-\mathcal{G}_{j,j+1}\mathcal{G}_{j+1,n}\mathcal{G}_{m,j}
				+\mathcal{G}_{j,j+1}\mathcal{G}_{m,j}\,\delta_{j+1,n}
				-\mathcal{G}_{j,n}\mathcal{G}_{j+1,j}\mathcal{G}_{m,j+1}\\
				&\quad
				+\mathcal{G}_{j,n}\mathcal{G}_{j+1,j}\,\delta_{j+1,m}
				+\mathcal{G}_{j,n}\mathcal{G}_{j+1,j+1}\mathcal{G}_{m,j}
				-\mathcal{G}_{j,n}\mathcal{G}_{m,j},
			\end{aligned}
		\end{equation}
	and
	\begin{equation}
		\begin{aligned}
			D_R
			=
			\langle
			\hat{c}_j^\dagger \hat{c}_{j+1}
			\hat{c}_{j+1}^\dagger \hat{c}_j
			\rangle
			\approx
			-\mathcal{G}_{j,j}\mathcal{G}_{j+1,j+1}
			+\mathcal{G}_{j,j}
			+\mathcal{G}_{j,j+1}\mathcal{G}_{j+1,j}.
		\end{aligned}
	\end{equation}
	The corresponding formulas for a left jump are obtained by exchanging $j\leftrightarrow j+1$.
	
	Within the Gaussian approximation, the nonunitary evolution generated by $\hat{H}_\mathrm{eff}$ can be written as
	\begin{equation}
		\dot{\mathcal{G}}_{mn}
		\approx
		\dot{\mathcal{G}}_{mn}^{(H)}
		+
		\dot{\mathcal{G}}_{mn}^{(\mathrm{lin})}
		+
		\dot{\mathcal{G}}_{mn}^{(\mathrm{quart})},
	\end{equation}
	with
		\begin{equation}
			\begin{gathered}
				\dot{\mathcal{G}}_{mn}^{(\mathrm{lin})}
				=
				\sum_{k=1}^L \mu_k \mathcal{G}_{mk}\mathcal{G}_{kn},\\
				\dot{\mathcal{G}}_{mn}^{(\mathrm{quart})}
				=
				J\sum_{j=1}^{L-1}
				\left[
				\left(
				\mathcal{G}_{j+1,j}\mathcal{G}_{j,n}
				-
				\mathcal{G}_{j,j}\mathcal{G}_{j+1,n}
				\right)\mathcal{G}_{m,j+1}
				+
				\left(
				\mathcal{G}_{j,j+1}\mathcal{G}_{j+1,n}
				-
				\mathcal{G}_{j+1,j+1}\mathcal{G}_{j,n}
				\right)\mathcal{G}_{m,j}
				\right],
			\end{gathered}
		\end{equation}
	where $\mu_1=(J-\gamma)/2$, $\mu_L=(J+\gamma)/2$, and $\mu_k=J$ for $k=2,\dots,L-1$. %
	For a quadratic Hamiltonian $\hat H=\sum_{mn} h_{mn}\hat c_m^\dagger \hat c_n$, the Hamiltonian contribution is
    \begin{equation}
        \dot{\mathcal{G}}^{(H)} = -i [\mathcal{G},h],
    \end{equation}
    with $h_{mn}$ given by Eq.~\eqref{eq:hamiltonian-power-law}.

	\section{Validity of the Gaussian approximation}
	\label{app:gaussian-approximation}
	
\begin{figure*}[tb]
	    \centering
        \includegraphics[width = \linewidth]{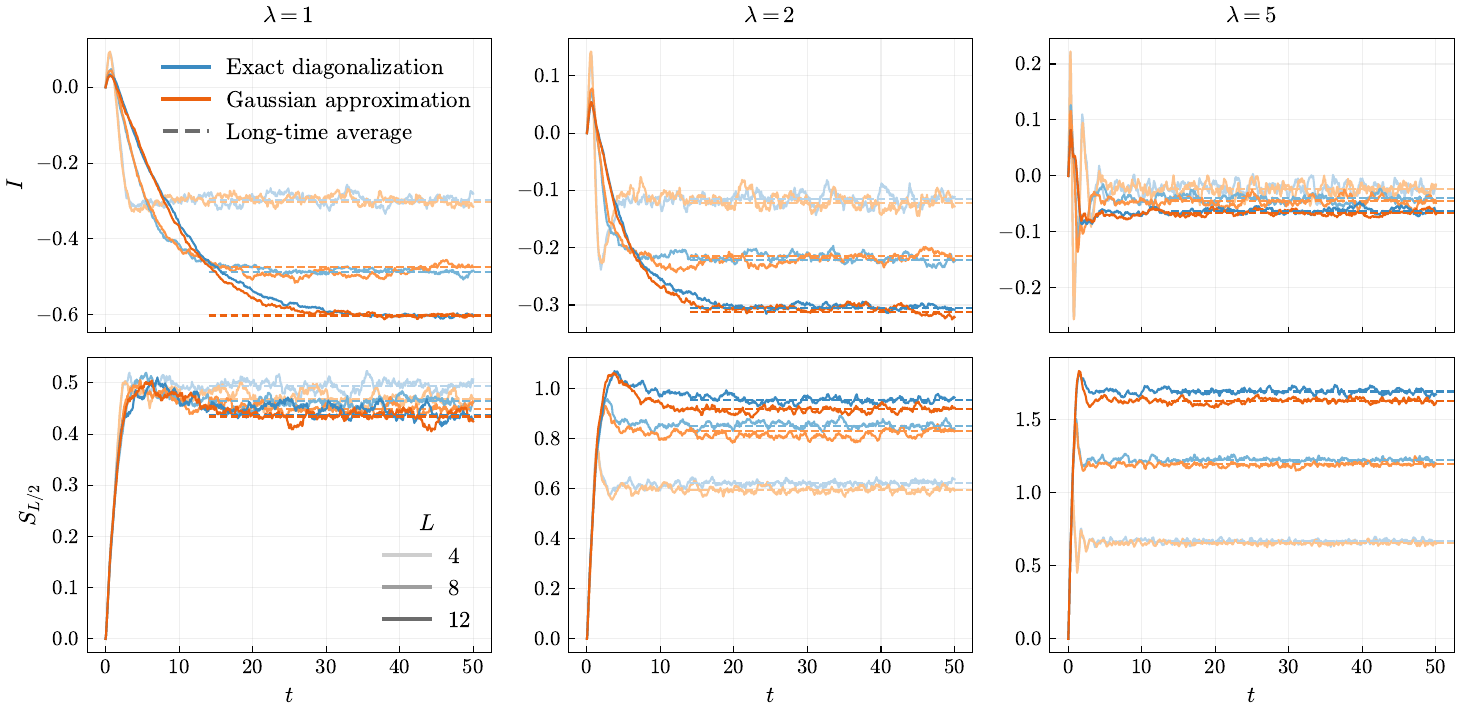}
        \caption{Comparison between exact and Gaussian dynamics of the imbalance (top row) and of the entanglement entropy (bottom row), for different values of the unitary coupling $\lambda$ in the nearest-neighbor Hamiltonian ($\gamma = 1/2$), and different system sizes. Blue curves are results of exact diagonalization, while red curves correspond to the Gaussian approximation. Darker curves correspond to larger system sizes. The dashed lines denote the long-time averages of the studied quantities.}
        \label{fig:gaussian-benchmark}
	\end{figure*}

	In this section we benchmark the Gaussian approximation used in the main text against exact trajectory simulations at small system sizes, where the full many-body dynamics can be followed exactly. First, we focus on the nearest-neighbor Hamiltonian with $\alpha\to\infty$. In Fig.~\ref{fig:gaussian-benchmark}, we compare the exact time evolution of the imbalance (top row) and entanglement entropy (bottom row) with that obtained from the Gaussian approximation, for different system sizes (color shading)  and different values of the unitary coupling $\lambda$ (columns), with $\gamma = 1/2$. Results are averaged over $N_T = 1000$ quantum trajectories. In all cases, the Gaussian approximation correctly reproduces the transient regime and the stationary values (dashed lines), and provides a good estimate of the scaling with the system size $L$.
	
\begin{figure*}[tb]
	    \centering
        \includegraphics[width = \linewidth]{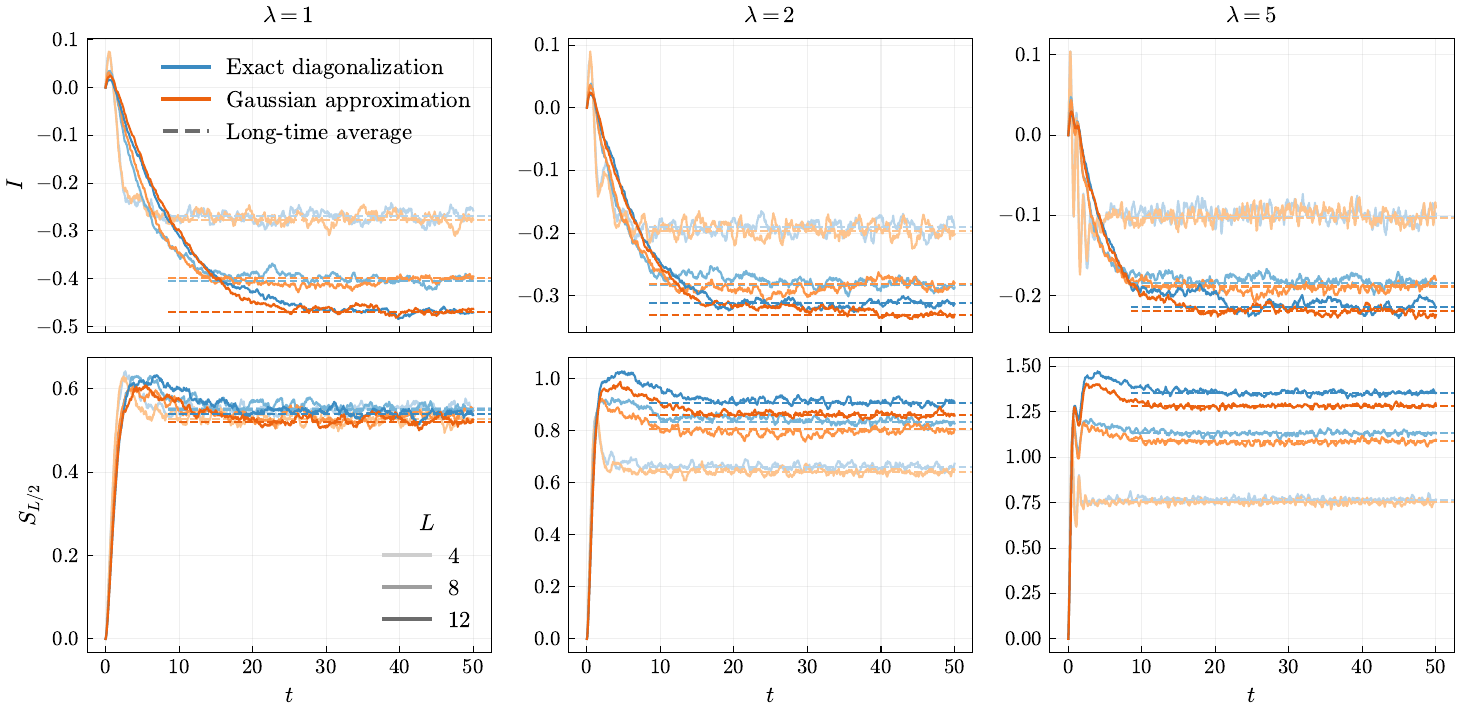}
        \caption{Same as Fig.~\ref{fig:gaussian-benchmark} but for the long-range model with $\alpha = 1/2$.}
        \label{fig:gaussian-benchmark-alpha-0.50}
	\end{figure*}

    An analogous benchmark is shown in Fig.~\ref{fig:gaussian-benchmark-alpha-0.50} for $\alpha = 1/2$. Here we can see that, while the Gaussian dynamics remains qualitatively similar to the exact diagonalization, the Gaussian approximation systematically underestimates the steady-state entanglement entropy of the ensemble, with deviations noticeably increasing with $L$ at large $\lambda$.
    This indicates that the Gaussian scheme is conservative in the long-range regime: the algebraic growth observed within this approximation is not an artifact of artificially enhanced entanglement. However, the exact asymptotic exponent cannot be reliably extracted from these small-size benchmarks.    

    \section*{Steady-state density profiles}

    This section provides additional information about the steady-state density profiles for different values of the hopping range $\alpha$. The density profile is an ensemble-averaged one-body observable, whereas the trajectory entanglement discussed in the main text is nonlinear and is not determined by the density profile alone.  They document how the real-space skin profile changes with $\alpha$ and provide an independent density-side view of the same range-controlled phenomenology. 
    
\begin{figure}[tb]
        \centering
        \includegraphics[width = \linewidth]{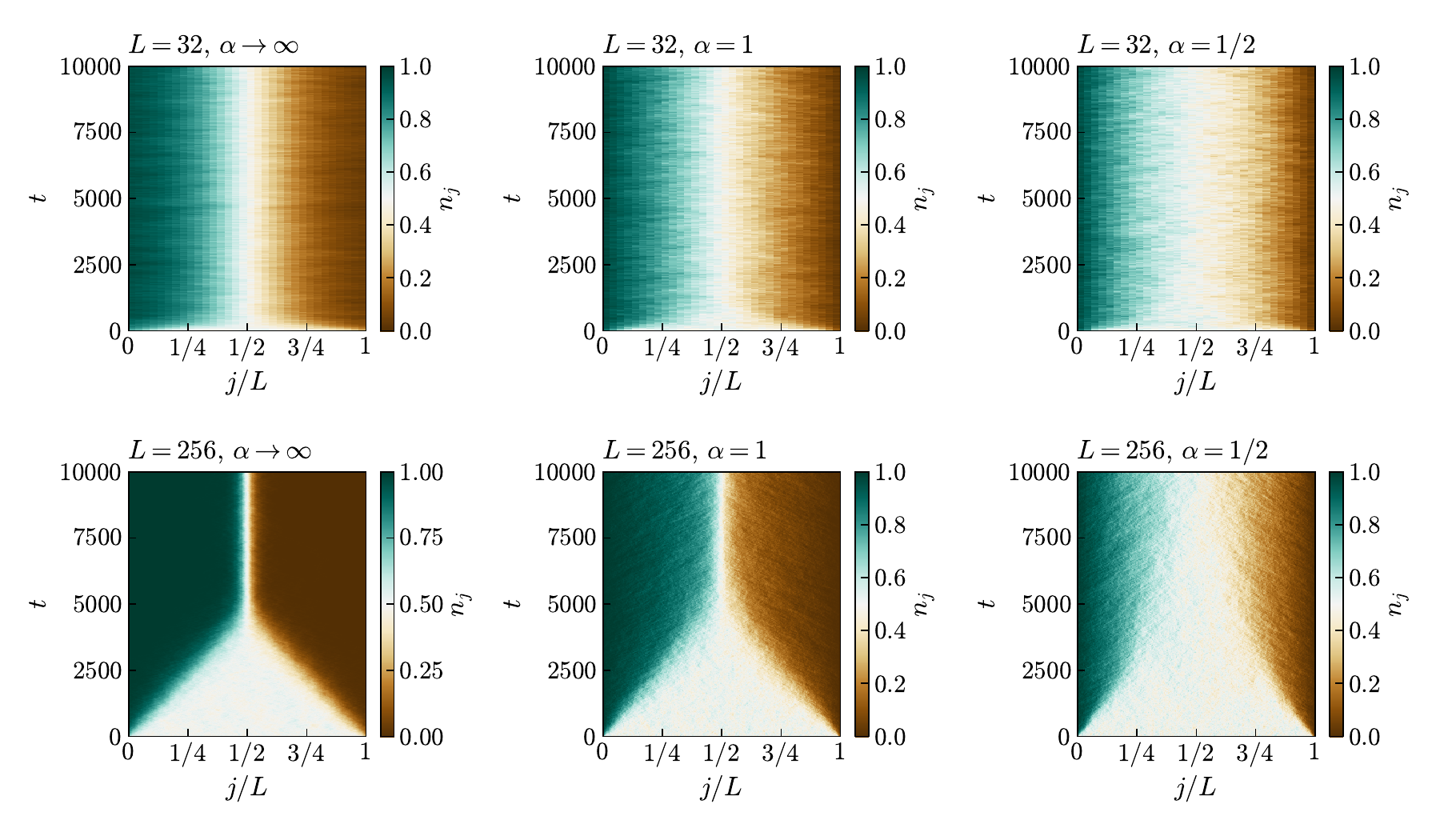}
        \caption{Density profiles as a function of lattice position and time for three representative hopping ranges and two system sizes ($\lambda=2$, $\gamma=1/2$). The short-range model develops a sharp Fermi-skin front whose width decreases relative to the system size. The marginal case $\alpha=1$ shows a broader interface and stronger finite-size effects. For $\alpha=1/2$, the right-half density tail persists at large $L$, consistently with incomplete skin accumulation.}
        \label{fig:heatmaps}
    \end{figure}

\begin{figure}[tb]
        \centering
        \includegraphics[width = \linewidth]{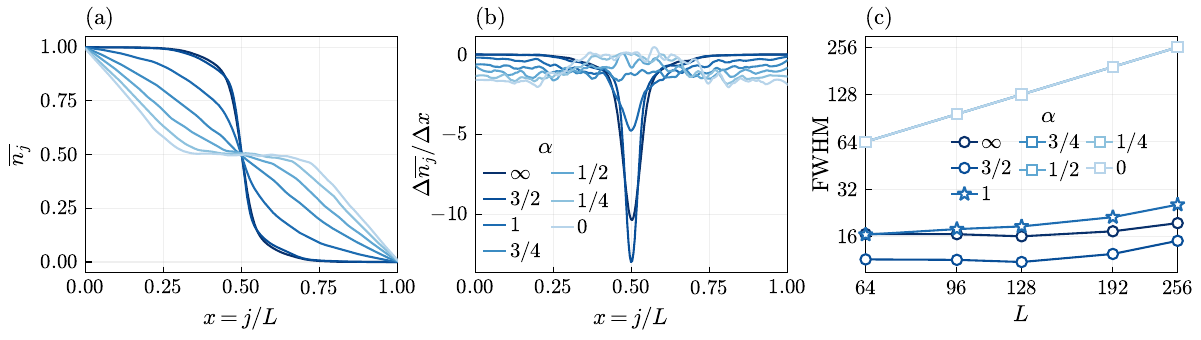}
        \caption{(a) Steady-state density profiles and (b) their numerical derivative as a function of lattice position, for different values of $\alpha$ ($L = 256$, $\lambda = 2$). (c) Full width at half maximum of the numerical derivatives of $\overline{n(x)}$, with $x = j/L$, as a function of $L$ ($\lambda = 2$). This quantity characterizes the density-profile crossover.}
        \label{fig:density}
    \end{figure}

    Figure~\ref{fig:heatmaps} shows the approach to the stationary density profile for representative ranges and sizes. In the short-range case, the density rapidly forms a sharp Fermi-skin front close to the center of the chain, and increasing $L$ mainly sharpens this front without producing a sizable right-half tail. For $\alpha=1/2$, instead, the stationary profile remains extended across a macroscopic portion of the right half, already at the largest size shown. The marginal case $\alpha=1$ displays an intermediate behavior, with a visibly broader interface and stronger finite-size dependence than the short-range profile. Figure~\ref{fig:density}(a) shows steady-state profiles for several values of $\alpha$ at $L=256$ and $\lambda=2$, while Fig.~\ref{fig:density}(b) shows the corresponding numerical derivative with respect to $x=j/L$. As $\alpha$ is decreased, the profile evolves from a sharp Fermi-skin step to an extended right-half tail. We quantify this density crossover by the full width at half maximum (FWHM) of the numerical derivative shown in Fig.~\ref{fig:density}(c). For the available sizes, the FWHM grows approximately extensively for $\alpha<1$, while for $\alpha>1$ it is much less size dependent, with the marginal region near $\alpha\simeq1$ showing the strongest finite-size effects. This supports the statement that $\alpha\simeq1$ is special for the density profile, whereas the entanglement crossover extracted from $\beta(\alpha)$ extends across the interval $1 \lesssim \alpha \lesssim 2$.

	\section{Additional results for \texorpdfstring{$\alpha = 3/2$}{alpha = 3/2} and \texorpdfstring{$\alpha = 1$}{alpha = 1}}
	
\begin{figure}[tb]
		\centering
		\includegraphics[width = \textwidth]{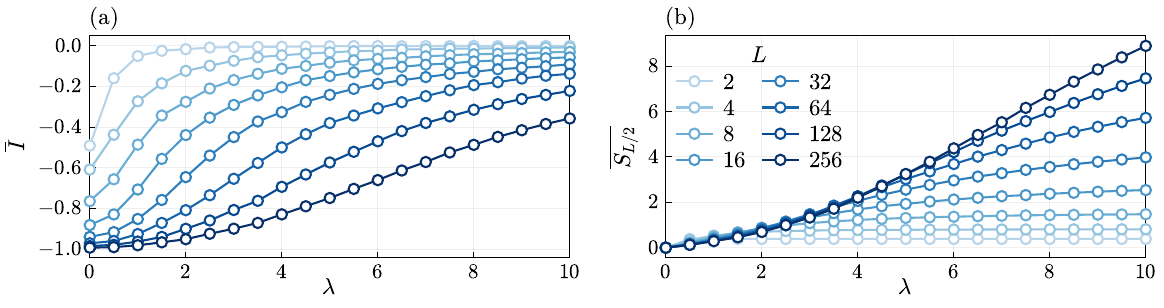}
		\caption{Steady-state properties for the intermediate-range Hamiltonian with $\alpha=3/2$ ($\gamma=1/2$). (a) Imbalance $\overline I$ and (b) half-chain entanglement entropy $\overline{S_{L/2}}$ as functions of the coherent coupling $\lambda$.}
		\label{fig:gamma-0.5-alpha-1.50}
	\end{figure}

	In Fig.~\ref{fig:gamma-0.5-alpha-1.50}, we show the steady-state imbalance and half-chain entanglement entropy for the intermediate-range model with $\alpha = 3/2$ ($\gamma = 1/2$). This value lies in the crossover window identified by the $\beta(\alpha)$ analysis. The imbalance flows toward stronger skin accumulation with increasing $L$, and the finite-size difference in Fig.~\ref{fig:gamma-0.5-nn-delta-entanglement}(b) shows that the onset of entanglement growth drifts to larger $\lambda$ over the accessible sizes. We therefore interpret $\alpha=3/2$ as an intermediate regime approaching the area-law side of the crossover, rather than as a sharp short-range boundary.

	\begin{figure}[tb]
		\centering
		\includegraphics[width = \textwidth]{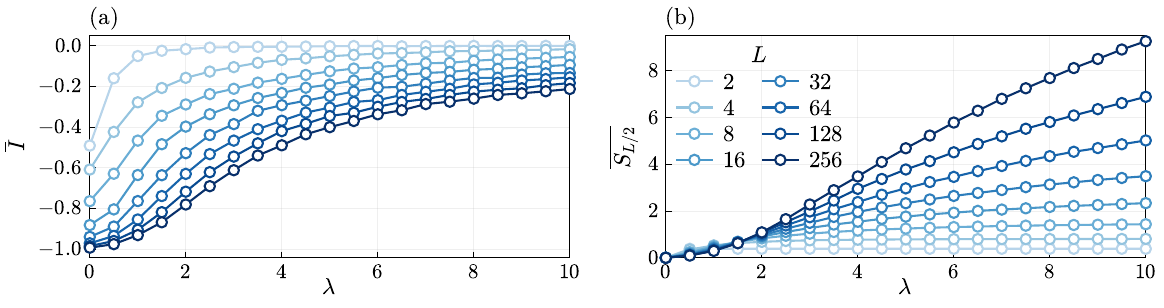}
		\caption{Steady-state properties for the Hamiltonian with $\alpha=1$ ($\gamma=1/2$). (a) Imbalance $\overline I$ and (b) half-chain entanglement entropy $\overline{S_{L/2}}$ as functions of the coherent coupling $\lambda$.}
		\label{fig:gamma-0.5-alpha-1.00}
	\end{figure}
	
	In Fig.~\ref{fig:gamma-0.5-alpha-1.00}, we show the steady-state imbalance and half-chain entanglement entropy for the model with $\alpha = 1$ ($\gamma = 1/2$). For this $\alpha$, the imbalance does not flow toward the fully imbalanced value in the accessible size window, but approaches a nonmaximal value, consistently with an incomplete LSE. The entanglement entropy grows with $L$, while the crossover point $\lambda_c$ between area-law and extensive scaling does not, as shown in Fig.~\ref{fig:gamma-0.5-nn-delta-entanglement}(c). Thus, in the thermodynamic limit, the system has sub-volume-law entanglement entropy for large values of $\lambda$. Since $\alpha = 1$ is marginal in the Kac-normalized long-range coupling, logarithmic finite-size corrections cannot be excluded.
	
	\section{Steady-state imbalance and entanglement versus system size}

\begin{figure*}[tb]
		\centering
		\includegraphics[width=\textwidth]{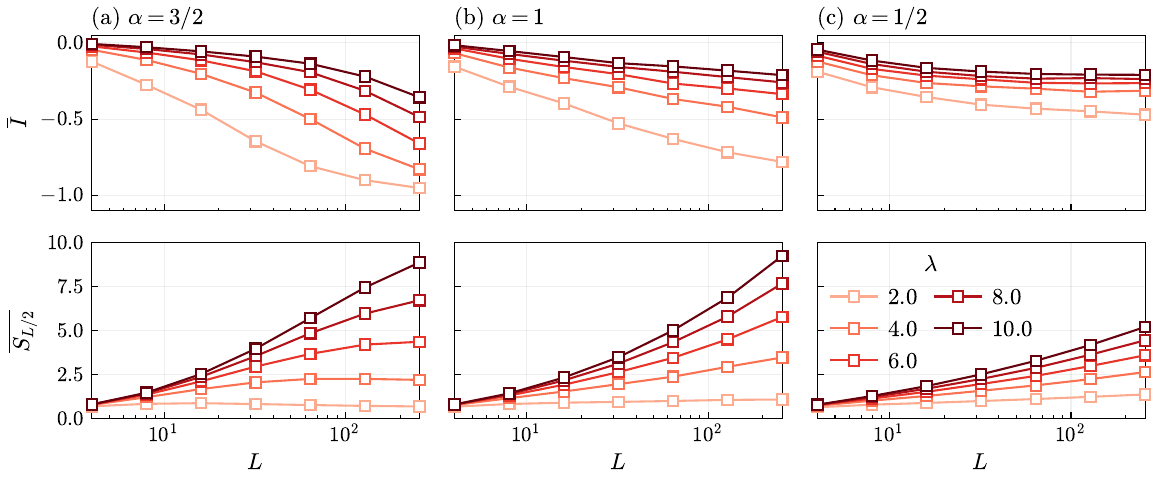}
		\caption{Steady-state imbalance (top row) and Gaussian trajectory half-chain entanglement entropy (bottom row) as functions of $L$, for different $\lambda$ and hopping ranges. (a) $\alpha=3/2$, (b) $\alpha=1$, and (c) $\alpha=1/2$. The imbalance data show a drift toward stronger skin accumulation for $\alpha=3/2$, while remaining nonmaximal for $\alpha=1$ and $\alpha=1/2$ over the accessible sizes. The entanglement panels show weak finite-size growth for $\alpha=3/2$ and more pronounced growth for $\alpha\le1$, consistently with the effective finite-size exponents discussed in the main text.}
		\label{fig:imbalance-vs-L}
	\end{figure*}

    In Fig.~\ref{fig:imbalance-vs-L}, we show the steady-state imbalance and half-chain entanglement entropy as functions of the system size, for different coherent couplings $\lambda$ and three hopping ranges ($J=1,\gamma=1/2$). The top row complements the imbalance analysis of the main text. For $\alpha=3/2$, the imbalance flows toward stronger skin accumulation with increasing $L$, although more slowly than in the nearest-neighbor limit. For $\alpha=1$ and $\alpha=1/2$, instead, the curves approach nonmaximal values over the accessible sizes, consistently with incomplete skin effect.

    The bottom row shows the corresponding Gaussian trajectory entanglement. For ($\alpha=3/2$, $S_{L/2}$ grows with $L$ at large $\lambda$, but the growth saturates at large $L$ and should be interpreted together with the finite-size difference in Fig.\ref{fig:gamma-0.5-nn-delta-entanglement}(b), where the onset shifts to larger $\lambda$ as $L$ increases. This is consistent with $\alpha=3/2$ lying on the area-law side of the crossover within the accessible sizes. For $\alpha=1$ and especially for $\alpha=1/2$, the entropy shows a more robust increase with $L$ at large $\lambda$.
    
	\section{Fitting parameters}
	
\begin{figure*}
        \centering
        \includegraphics[width=\textwidth]{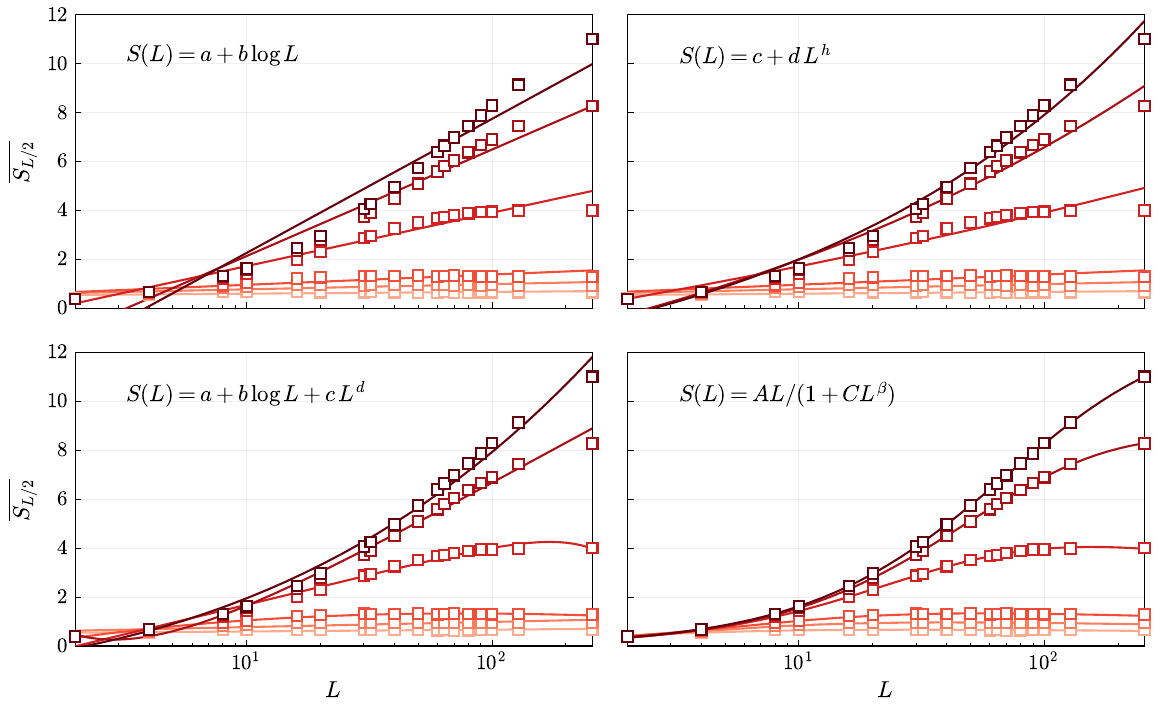}
        \caption{Numerical data fitting using different scaling relations for the entanglement entropy as a function of the system size. Eq.~\eqref{eq:entanglement-fit-s} yields the most accurate results in all the analyzed parameter regimes.
        Here we consider a nearest-neighbor Hamiltonian and fix $\gamma=1/2$.}
        \label{fig:fit-accuracy}
    \end{figure*}

	To infer the behavior of the half-chain entanglement entropy at large sizes, several fitting functions can be used. In our analysis, we tested the following laws:
    \begin{subequations}
       \begin{eqnarray}
          S(L) & = & a + b \log L,\\
          S(L) & = & c + d \, L^h,\\
          S(L) & = & a + b \log L + c \, L^d.
       \end{eqnarray}
    \end{subequations}
    These forms give visibly large residuals over the full range of sizes and coherent couplings analyzed. On the other hand, the empirical fitting law 
	\begin{equation}
		\label{eq:entanglement-fit-s}
		S(L)=\frac{A L}{1+C L^\beta} \,, \qquad A,C,\beta\ge 0 \,,
	\end{equation}
	works remarkably well. The comparison between the different fitting functions is shown in Fig.~\ref{fig:fit-accuracy} for the nearest-neighbor case ($\gamma = 1/2$). As mentioned in the main text, Eq.~\eqref{eq:entanglement-fit-s} captures an initial linear growth crossing over to saturation in the presence of dissipation, and serves as a diagnostic of the scaling regime: $\beta=0$ corresponds to volume-law scaling, $0<\beta<1$ to sub-volume-law behavior, and $\beta\ge 1$ to area-law scaling, with the marginal case $\beta=1$ allowing for possible logarithmic corrections. 

\begin{figure*}
		\centering
		\includegraphics[width=\textwidth]{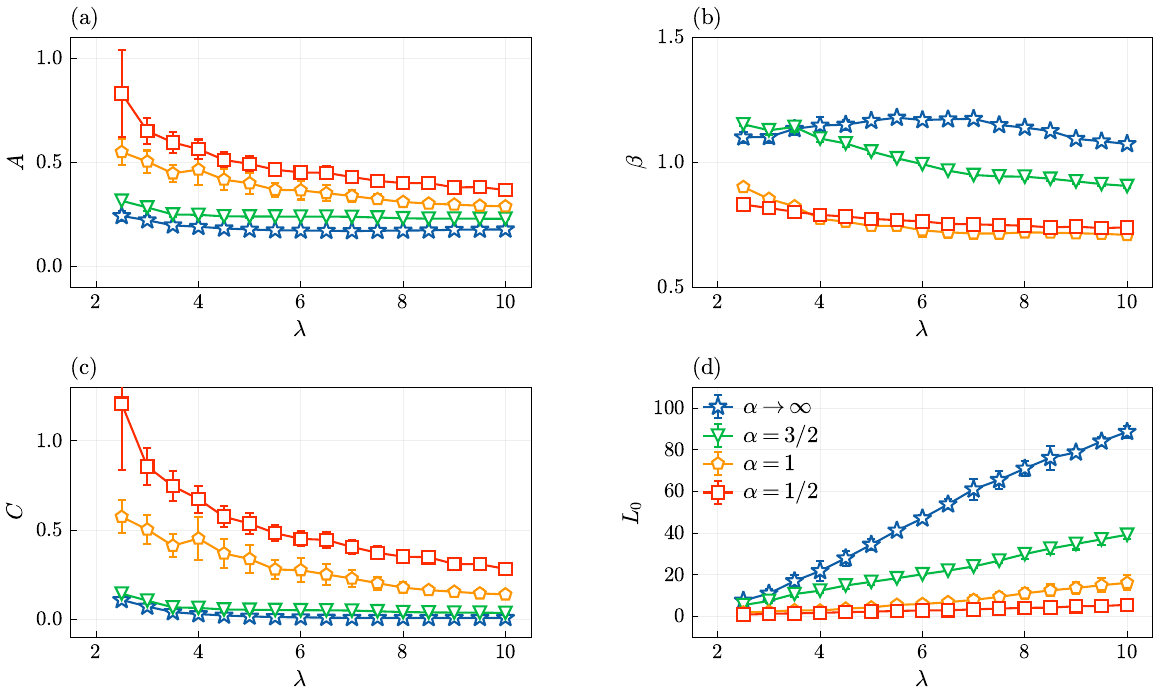}
		\caption{Results of the fit to Eq.~\eqref{eq:entanglement-fit-s} for all the analyzed models.}
		\label{fig:entanglement-vs-L-fitting-parameters}
	\end{figure*}

    The fit is meaningful only when the available sizes extend beyond the characteristic scale $L_0=C^{-1/\beta}$. Figure~\ref{fig:entanglement-vs-L-fitting-parameters} summarizes the fitting results for the models analyzed in the main text and in this Supplemental Material. The fit remains reliable for all analyzed $\lambda$, since $L_0 < 100$. Panel~\ref{fig:entanglement-vs-L-fitting-parameters}(b) shows that the hopping range strongly affects the scaling behavior of the entanglement entropy.

    \section{Cases without skin: \texorpdfstring{$\gamma = 0$}{gamma = 0} and periodic boundary conditions}

    To highlight the role of the Fermi-skin domain in the phenomenology discussed so far, we consider here cases in which it is absent. Our purpose is to separate the effect of long-range hopping alone from the combined effect of long-range hopping and directed dissipative accumulation. In the presence of a skin domain, short-range hopping yields area-law steady-state entanglement, whereas sufficiently long-range hopping couples the skin region to the bulk and produces algebraically growing entanglement entropy. In the absence of asymmetric accumulation, the scenario changes qualitatively: the hopping range no longer controls the drift of the finite-size crossover, indicating that the skin domain is essential for the range-controlled phenomenology discussed in the main text.

\begin{figure}
        \centering
        \includegraphics[width=\linewidth]{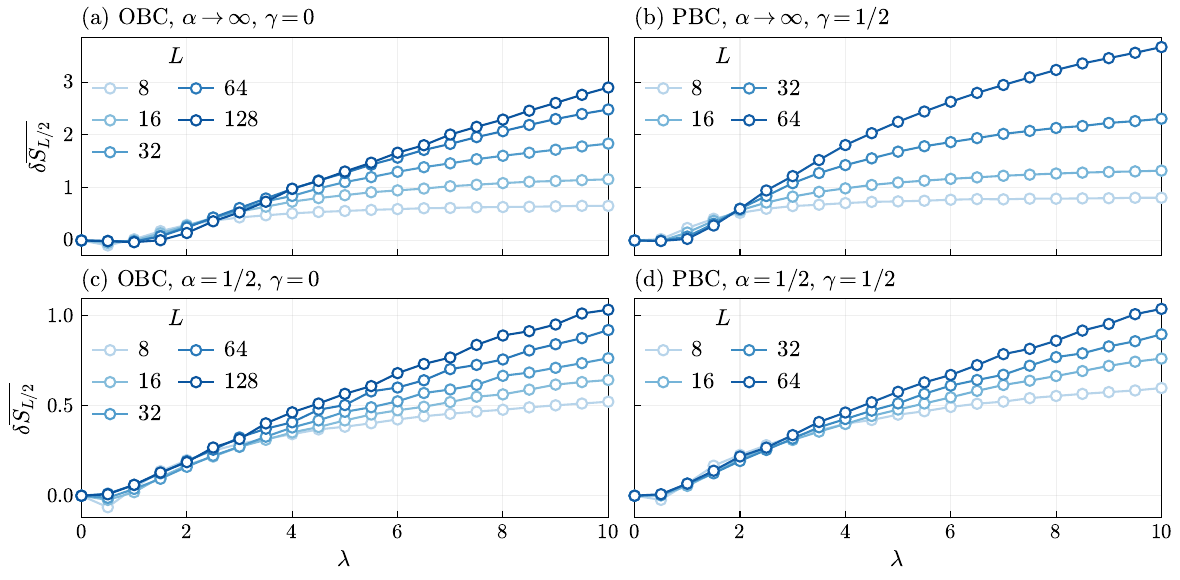}
        \caption{Finite-size difference $\delta \overline{S_{L/2}}$ in cases without a skin domain. 
        (a-c) Open boundary conditions with no asymmetry ($\gamma = 0$), for $\alpha\to\infty$ and $\alpha = 1/2$, respectively. 
        (b-d) Periodic boundary conditions with asymmetry ($\gamma = 1/2$), for $\alpha\to\infty$ and $\alpha = 1/2$, respectively. In all cases, the onset of entanglement growth remains approximately size-independent.}
        \label{fig:dS-noskin}
    \end{figure}

    To show this, we first consider $\gamma = 0$. In this case, the left-right dissipation asymmetry is absent and the steady state exhibits a uniform density profile for all values of $\alpha$ and $\lambda$: the steady-state left-right imbalance therefore vanishes identically. On the other hand, the monitored ensemble of quantum trajectories still develops entanglement, but its dependence on the system size $L$ is markedly different from the asymmetric case with $\gamma = 1/2$. In Fig.~\ref{fig:dS-noskin}(a,c) we report the finite-size difference $\delta \overline{S_{L/2}}$ as a function of $\lambda$ for the nearest-neighbor model ($\alpha\to\infty$, panel a) and the long-range model ($\alpha = 1/2$, panel c) in the symmetric case. In both cases, we observe a crossover between area-law and sub-volume-law behavior. However, for the nearest-neighbor model the crossover does not drift with the system size, and the system remains sub-volume-law entangled in the thermodynamic limit at large $\lambda$. This can be understood as follows: in the absence of a skin domain, the system remains delocalized as $L\to \infty$, therefore the mechanism responsible for the shift of the crossover point to larger and larger $\lambda$ in the asymmetric open-boundary case is no longer active.

    Another way to eliminate the skin domain is to impose periodic boundary conditions. In this case, even for $\gamma \ne0$ particles cannot accumulate due to the absence of boundaries, and the steady-state imbalance remains identically zero. Here we consider $\gamma = 1/2$ and show in Fig.~\ref{fig:dS-noskin}(b,d) the finite-size difference $\delta \overline{S_{L/2}}$ as a function of $\lambda$ for the nearest-neighbor model ($\alpha\to\infty$, panel b) and the long-range model ($\alpha = 1/2$, panel d). Also in this case, the data are consistent with a size-independent crossover between area-law and algebraic entanglement growth, independently of the hopping range. Thus, the existence of the skin domain is crucial in shaping the entanglement structure of the steady state: without it, hopping range alone does not reproduce the phenomenology of the asymmetric open chain.
    
\end{document}